\documentclass[onecolumn]{mem}
\usepackage{natbib}\usepackage{txfonts}\usepackage{balance}
\bibpunct{(}{)}{;}{a}{}{,} 
\usepackage{graphicx}
\usepackage[a4paper]{hyperref}
\idline{0}{0}  
\begin{document}
\def\teff{$T\rm_{eff }$}
\def\kms{$\mathrm {km s}^{-1}$}

\def\pn{\par\noindent}
\def\bs{\bigskip}
\def\ms{\medskip}
\def\ss{\smallskip}
\newcommand{\ls}{{_<\atop^{\sim}}}
\newcommand{\lax}{{_<\atop^{\sim}}}
\newcommand{\gs}{{_>\atop^{\sim}}}
\newcommand{\gax}{{_>\atop^{\sim}}}


\arraycolsep0.35mm                      
\catcode`\@=11
\def\gsim{\,\mathrel{\mathpalette\@versim>\,}}
\def\lsim{\,\mathrel{\mathpalette\@versim<\,}}
\def\@versim#1#2{\lower 2.9truept \vbox{\baselineskip 0pt \lineskip
    0.5truept \ialign{$\m@th#1\hfil##\hfil$\crcr#2\crcr\sim\crcr}}}
\catcode`\@=12  
\newcommand{\az}{a_0}
\newcommand{\gz}{g_0}
\newcommand{\gv}{{\bf g}}
\newcommand{\gtv}{{\tilde\gv}}
\newcommand{\gt}{{\tilde{g}}}
\newcommand{\dgv}{\delta{\bf g}}
\newcommand{\xv}{{\bf x}}
\newcommand{\de}{{\rm d}}
\newcommand{\rhoz}{\rho_0}
\newcommand{\varrhoz}{\varrho_0}
\newcommand{\phiz}{\phi_0}
\newcommand{\phis}{\phi_{\rm s}}
\newcommand{\mus}{\mu_{\rm s}}
\newcommand{\phiini}{\phi^{(0)}}
\newcommand{\phizgamma}{\phi_{0,\gamma}}
\newcommand{\phiu}{\phi_1}
\newcommand{\Mz}{M_0}
\newcommand{\tmax}{t_{\rm max}}
\newcommand{\Smath}{{\mathcal{S}}}
\newcommand{\Smathlm}{{\mathcal{S}}_{l,m}}
\newcommand{\Smathlmnow}{{\mathcal{S}}_{l,m}^{(n)}}
\newcommand{\rhoNz}{\rho_{\rm N0}}
\newcommand{\rhoMz}{\rho_{\rm M0}}
\newcommand{\rhou}{\rho_{\rm 1}}
\newcommand{\Mu}{M_{\rm u}}
\newcommand{\lu}{l_{\rm u}}
\newcommand{\tu}{t_{\rm u}}
\newcommand{\vu}{v_{\rm u}}
\newcommand{\au}{a_{\rm u}}
\newcommand{\rhoNu}{\rho_{\rm N1}}
\newcommand{\phiN}{\phi_{\rm N}}
\newcommand{\psiu}{\psi_1}
\newcommand{\psid}{\psi_2}
\newcommand{\Epot}{E_{\rm pot}}
\newcommand{\gvnum}{{\bf g}_{\rm num}}
\newcommand{\phinum}{\phi_{\rm num}}
\newcommand{\gvN}{\gv_{\rm N}}
\newcommand{\gN}{g_{\rm N}}
\def\dxcube{d^3{\bf x}}
\newcommand{\hv}{{\bf h}}
\newcommand{\Svu}{{\bf S}_1}
\newcommand{\Sv}{{\bf S}}
\newcommand{\rc}{r_{\rm c}}    
\newcommand{\sthsq}{\sin^2\vartheta}
\newcommand{\cthsq}{\cos^2\vartheta}
\newcommand{\sphsq}{\sin^2\varphi}
\newcommand{\sth}{\sin\vartheta}
\newcommand{\cth}{\cos\vartheta}
\newcommand{\sph}{\sin\varphi}
\newcommand{\etatilde}{\tilde{\eta}}
\newcommand{\epsilontilde}{\tilde{\epsilon}}
\renewcommand{\baselinestretch}{1.2}
\newcommand{\be}{\begin{equation}}
\newcommand{\ee}{\end{equation}}
\newcommand{\p}{\partial}
\def\parn {\par\noindent}
\def\vv {{\bf v}}
\def\xx {{\bf x}}
\def\xp {x_p}
\def\yp {y_p}
\def\zp {z_p}
\def\rp {r_p}
\def\xip {\xi_p}
\def\GG{{\bf G}}
\def\dGG{\delta\GG}
\def\grp{\nabla\phi}
\def\pz{\phi^{(0)}}
\def\gradpz{\nabla\pz}
\def\omeganow{\omega}
\def\phinow{\phi^{(n)}}
\def\phinowlm{\phi_{l,m}^{(n)}}
\def\phiplus{\phi^{(n+1)}}
\def\phiminus{\phi^{(n-1)}}
\def\Gn{\GG^{(n)}}
\def\gvnow{\gv^{(n)}}
\def\gnow{g^{(n)}}
\def\gplus{g^{(n+1)}}
\def\gvplus{\gv^{(n+1)}}
\def\dGn{\delta\Gn}
\def\dgvnow{\delta {\bf g}^{(n)}}
\def\dphi{\delta\phi}
\def\dphinow{\delta\phinow}
\def\dphiplus{\delta\phiplus}
\def\dphiminus{\delta\phiminus}
\def\dphinowlm{\delta\phinowlm}
\def\dtx{d^3\xx}
\def\Nxi{N_{\xi}}
\def\Nr{N_{r}}
\def\Nth{N_{\vartheta}}
\def\Npe{N_{\rm PE}}
\def\Nph{N_{\varphi}}
\def\Dxi{\Delta\xi}
\def\Dr{r^{\prime}(\Sxi)\Dxi}
\def\Dth{\Delta\vartheta}
\def\Dph{\Delta\varphi}
\def\En{{\cal E}}
\def\Ra{R_{\rm a}}
\def\rhotil{\tilde\rho}
\def\vv{{\bf v}}
\def\dtv{d^3\vv}
\def\hL {\hat{L}}
\def\hM {\hat{M}}
\def\Msun{M_{\odot}}
\def\kpc{{\rm kpc}}
\def\yr{{\rm yr}}
\def\kms{{\rm km\,s^{-1}}}  
\def\dhM{\delta\hat{M}}
\def\dhMnow{\delta\hat{M}^{(n)}}
\def\dhMminus{\delta\hat{M}^{(n-1)}}
\def\Rnow{\hat{R}^{(n)}}
\def\dhMbar{\delta\hat{\mathcal{M}}^{(n)}}
\def\mubar{\bar{\mu}^{(n)}}
\def\munow{\mu^{(n)}}
\def\muprimenow{\mu^{\prime(n)}}
\def\muplus{\mu^{(n+1)}}
\def\muminus{\mu^{(n-1)}}
\def\hf {\hat{f}}
\def\pph {\phi}
\def\phin{\pph_n}
\def\phinm{\pph_{n-1}}
\def\phinp{\pph_{n+1}}
\def\dphi{\delta\pph}
\def\dphin{\delta\phin}
\def\dphinm{\delta\phinm}
\def\linop{\mathcal{L}}
\def\II {\bf{I}}
\def\FF {\bf{F}}
\def\JJ {\hat{J}}

\title{N-MODY: a code for collisionless N-body simulations
in modified Newtonian dynamics}

   \subtitle{}

\author{
P. \,Londrillo\inst{1} 
\and C. \, Nipoti\inst{2}
          }
  \offprints{P. Londrillo}

\institute{
Istituto Nazionale di Astrofisica --
Osservatorio Astronomico di Bologna, Via Ranzani 1,
I-40127 Bologna, Italy
\email{pasquale.londrillo@oabo.inaf.it}
\and
Dipartimento di Astronomia -- Universit\`a di Bologna, via Ranzani 1,
I-40127 Bologna, Italy
\email{carlo.nipoti@unibo.it}
}

\authorrunning{Londrillo \& Nipoti}

\titlerunning{N-body simulations in MOND}

\abstract{We describe the numerical code N-MODY, a parallel
  particle-mesh code for collisionless N-body simulations in modified
  Newtonian dynamics (MOND). N-MODY is based on a numerical potential
  solver in spherical coordinates that solves the non-linear MOND
  field equation, and is ideally suited to simulate isolated stellar
  systems. N-MODY can be used also to compute the MOND potential of
  arbitrary static density distributions. A few applications of N-MODY
  indicate that some astrophysically relevant dynamical processes are
  profoundly different in MOND and in Newtonian gravity with dark
  matter.

\keywords{gravitation --- stellar dynamics --- galaxies: kinematics
   and dynamics  --- methods: numerical } } 

\maketitle{}

\section{Introduction}

Modified Newtonian dynamics (MOND) is an alternative gravity theory,
originally proposed by \cite{Mil83} to explain the observed kinematics
of disk galaxies without dark matter.  In Bekenstein \& Milgrom's
Lagrangian formulation of MOND \citep{Bek84} the Poisson equation is
replaced by the non-linear field equation
\begin{equation}
\nabla\cdot\left[\mu\left({\Vert\nabla\phi\Vert\over\az}\right)
\nabla\phi\right] = 4\pi G \rho,
\label{eqMOND}
\end{equation}
where $\az\simeq 1.2 \times 10^{-10} {\rm m\, s^{-2}}$ is a
characteristic acceleration, $\Vert ...\Vert$ is the standard
Euclidean norm in $\mathbb{R}^3$, and $\phi$ is the MOND gravitational
potential produced by the density distribution $\rho$.  The MOND
interpolating function $\mu(y)$ is a monotonic function such that
\begin{equation}
\mu(y)\sim\cases{y&for $y\ll 1$,\cr 1&for $y\gg 1$.}
\label{eqmulim}
\end{equation}
For finite mass systems the boundary condition of
equation~(\ref{eqMOND}) is $\nabla\phi\to 0$ for
$\Vert\xv\Vert\to\infty$.

As MOND is successful in reproducing several observed properties of
galaxies \citep[e.g.][]{Bek06}, there is growing interest in studying
dynamical processes in MOND with the aid of N-body simulations. While
in Newtonian gravity N-body codes can take advantage of the multipole
expansion of the Green function \citep[treecodes;][]{Bar86,Deh02}, the
non-linearity of the MOND field equation forces one to solve for the
potential on a grid, as done in particle-mesh schemes
\citep[see][]{Hoc88}.

We developed N-MODY, a parallel three-dimensional particle-mesh code
for collisionless N-body simulations in MOND.  The N-body code and the
potential solver on which the code is based have been already tested
and applied in \cite{Cio06,Cio07} and
\cite{Nip07a,Nip07b,Nip07c,Nip08}. The potential solver of N-MODY
solves the MOND field equation~(\ref{eqMOND}) using a relaxation
method in spherical coordinates based on the spherical harmonics
expansion. Thus the code is ideally suited for simulations of isolated
stellar systems \citep[but see][for an application to galaxy
merging]{Nip07c}. N-MODY can also be used to compute the MOND
potential of arbitrary static density distributions.  N-MODY is one of
the very few MOND N-body code developed so far: as far as we know the
only other three-dimensional MOND N-body code is Brada and Milgrom's
code \citep{Bra99}, which is based on a multi-grid potential solver in
Cartesian coordinates and has been recently implemented also by
\cite{Tir07}.

\section{Overview of the code}
\label{secove}

N-MODY implements a particle-mesh scheme in spherical coordinates,
following these main computational steps:
\begin{enumerate}
\item[1)] for a given distribution of $N$ particles, a grid-based
density field is reconstructed by mass deposition with linear or
quadratic shape functions.  Particles are represented by a
six-component array $(\xx_p,\vv_p)$ of positions and velocities in
Cartesian components ($p=1,...,N$).
\item[2)] For a given density distribution, the MOND acceleration and
potential fields are computed on the grid points. As an alternative,
the code also provides a fast Newtonian solver.
\item[3)] To move particles, the spherical components of the
acceleration are first interpolated at each particle position (using
the same linear or quadratic shape functions used in the mass
deposition) and then transformed into Cartesian components.
\item[4)] Finally, particle positions and velocities are advanced in
time using a leapfrog scheme of either second or fourth order.
\end{enumerate}

In N-MODY steps 1) and 3) are parallelized using MPI routines:
particles are uniformly distributed among the processors (PEs)
following the sequential ordering provided by their initial memory
addresses and then never exchanged among different PEs. This simple
strategy assures a full efficiency in parallel execution, but entails
more memory resources than in a standard domain decomposition. In
fact, grid data computed on the overall computational domain must be
at disposal of each PE at each timestep for particles interpolation
and move.  Step 2), which is only partially parallelized, contains the
MOND potential solver, which represents a new contribution to the fast
numerical solution of a non-linear elliptic equation. We now discuss
in more detail the different parts of the code: the potential solver
is described in Section~\ref{secsol}, while the particle-mesh scheme
and the time integration are described in Section~\ref{secpm}.

\section{The MOND potential solver}
\label{secsol}

N-MODY solves the non-relativistic MOND field
equation~(\ref{eqMOND}). By default in N-MODY we adopt as
interpolating function $\mu(y)={y/\sqrt{1+y^2}}$, but it is clear that
a trivial modification of the code allows to implement any other
continuous function $\mu$ with the required asymptotic
behaviour.\footnote{In versions of MOND based on Bekenstein's
covariant theory TeVeS \citep{Bek04}, there is a scalar field $\phis$
obeying equation~(\ref{eqMOND}), but with an unbounded interpolating
function $\mus$ instead of the bounded function $\mu$.  We verified
that our code can be adapted to solve for $\phis$
\cite[see][]{Fam07}. } N-MODY can also be used to simulate a Newtonian
system, in which case the Poisson equation is solved instead of
equation~(\ref{eqMOND}), or a system in the so-called `deep MOND
regime', i.e. obeying the equation
$\nabla\cdot\left({\Vert\nabla\phi\Vert}\nabla\phi\right) = 4\pi G \az
\rho$.

We consider only systems of finite mass $M=\int\rho(\xx)\,\dxcube$,
for which the boundary conditions of equation~(\ref{eqMOND}) are given
by $\Vert\nabla\phi\Vert\to O(1/r)$ for
$r\equiv\Vert\xv\Vert\to\infty$.  It must be stressed that the
asymptotic behaviour of the MOND potential ($\phi \sim \ln r \to
\infty$ for $r\to\infty$) is profoundly different from that of the
Newtonian potential ($\phi \to 0 $ for $r\to\infty$).  For this
reason, to solve numerically the MOND field equation (i) we discretize
a sufficiently large computational domain, in a way the asymptotic
boundary condition can be represented with reasonable accuracy and
(ii) we use a relaxation method to solve the non--linear elliptic
equation~(\ref{eqMOND}) with guess solution having the correct
asymptotic behaviour.

\subsection{The computational grid}

To accomplish task (i) above, N-MODY uses a spherical grid
$(r,\vartheta,\varphi)$ with radial coordinate represented by the
invertible mapping
\begin{equation}
r(\xi)=L\tan^{\alpha}\xi,\quad r^{\prime}(\xi)={\alpha L\tan^{\alpha-1}\xi\over\cos^2\xi},
\label{eqmapping}
\end{equation}
where $0\le\xi < \pi/2$, the mapping index $\alpha=1$ or $\alpha=2$
and the scale length $L$ are user provided parameters
\citep[see][]{Lon90}.  In this representation, the unbounded radial
range $(0,\infty)$ is mapped onto the finite open interval
$[0,\pi/2)$. The radial derivative is then expressed as
\begin{equation}
{\partial\over\partial r}={1\over r^{\prime}(\xi)}{\partial\over\partial\xi}
\end{equation}
and the $\xi$ coordinate is discretized into the uniform grid
\begin{equation}
\xi_i=(i+1/2)\Dxi,\quad\Dxi={\pi\over 2\Nr},\quad
i=0,1,..,\Nr,
\end{equation}
so the corresponding discretized radial variable $r_i=r(\xi_i)$ avoids
the singular points $r=0$ and $r=\infty$.  The other coordinates
$(\vartheta,\varphi)$ are discretized in the uniform grids
\begin{equation}
\vartheta_j=(j+1/2)\Delta_{\vartheta},\quad\Delta_{\vartheta}=\pi/\Nth,\quad
j=0,1,..,\Nth-1,
\end{equation}
to avoid the singular points $\vartheta=0$ and $\vartheta=\pi$, and
\begin{equation}
\varphi_k=k\Delta_{\varphi},\quad\Delta_{\varphi}=2\pi/\Nph,\quad
k=0,1,..,\Nph-1.
\end{equation}
The corresponding $(\xi,\vartheta,\varphi)$ numerical derivatives are
computed by second-order or fourth-order finite difference central
schemes.

\subsection{A Newton-like relaxation procedure}
\label{Newton}

 In N-MODY we use a Newton-like relaxation procedure to solve the
non--linear MOND equation, which can be written as
\begin{equation}
\hM[\phi(\xx)]\equiv\nabla\cdot\left[\mu\left({g\over\az}\right)
    \nabla\phi(\xx)\right]-4\pi G\rho(\xx)=0,\quad 
    g=O(r^{-1})\quad{\rm for }\quad r\to\infty,
\label{eqdiv}
\end{equation}
where $\rho(\xx)$ is assigned, $g(\xx)=\Vert\gv(\xx)\Vert$, with
$\gv(\xx)=-\nabla\phi(\xx)$, and $\xx=(r,\vartheta,\varphi)$.  Starting with
an approximate guess solution $\phiini$ having the required asymptotic
behaviour, a sequence of linear problems
\begin{equation}
\Rnow\dphinow=-\hM\left[\phinow\right],\quad n=0,1,...
\label{eqnew}
\end{equation} 
is solved for the increments $\dphinow=\phiplus-\phinow$, each
$\phinow$ provisional solution having the same asymptotic behaviour as
the guess $\phiini$.  Here the choice of the relaxation linear
operator $\Rnow$ is based on the requirement that it must assure
convergence in the maximum norm $\Vert ...\Vert_{\infty}$, i.e.
\begin{equation}
\left\Vert\dphinow\right\Vert_{\infty}=\left\Vert\left[\Rnow\right]^{-1}\hM\left[\phinow\right]\right\Vert_{\infty}
\le\left\Vert\dphiminus\right\Vert_{\infty},\quad n=1,2,...,
\label{eqconv}
\end{equation} 
and it must be easy to invert.  In a classical Newton method, one
would use $\Rnow=\dhMnow$, where the linear operator $\dhMnow$ is such
that
\begin{equation}
\dhMnow\left[\dphinow\right]= 
\hM\left[\phiplus\right]-\hM\left[\phinow\right]+O\left[(\dphinow)^2\right].
\end{equation}
For the specific case of $\hM$ defined by equation~(\ref{eqdiv}), 
\begin{equation}
\dhMnow=\munow\nabla^2 +\dhMnow_1,
\end{equation}
where
\begin{equation}
\dhMnow_1=(\nabla\munow)\cdot\nabla+\nabla\cdot\left[ \muprimenow\gvnow
{\left(\gvnow\cdot\nabla\right)} \right]
\label{eqdhM}
\end{equation}
and $\muprimenow\equiv \mu^{\prime}\left({\gnow/\az}\right)/\gnow\az$.
Boundedness of the inverse of the operator $\dhMnow$ assures quadratic
convergence of the scheme for $\phiini$ sufficiently close to the
sought solution \citep[e.g.][]{Sto80}. Unfortunately, $[\dhMnow]^{-1}$
is difficult to compute, so we discretize the simpler linear operator
\begin{equation}
\Rnow = \omega\munow\nabla ^2,
\label{eqomega}
\end{equation}
where $\omega> 1$ is an empirical relaxation parameter. As an
approximation of the Newton relaxation operator $\dhM$, this choice
assures a lower convergence rate, but has clear computational
advantages.  Using equation~(\ref{eqnew}) and the identity
\begin{equation}
\hM(\phinow)=\dhMminus\left[\dphiminus\right]+\hM\left[\phiminus\right]+O\left[(\dphiminus)^2\right], 
\end{equation}
it follows that the condition for convergence~(\ref{eqconv}) requires
\begin{equation}
\left\Vert{(\nabla^2)^{-1}\dhMminus\over\omega\munow}-{\muminus\over\munow}I\right\Vert_{\infty} < 1.
\end{equation}
Thus N-MODY solves the sequence of Poisson equations:
\begin{equation}
\nabla ^2\dphinow=\Smath^{(n)},\quad n=0,1,...,
\label{poisson}
\end{equation}
with source term given by 
\begin{equation}
\Smath^{(n)}=-{1\over\omega\munow}\hM\left[\phinow\right].
\end{equation}
In spherical coordinates, the Laplacian operator has the form
\begin{equation} 
\nabla^2\equiv{1\over r^2}\left[{\partial \over \partial
r}\left({r^2{\partial \over \partial r}}\right) 
+\hL_{\vartheta}+\hL_{\varphi}\right],
\label{eqLap}
\end{equation}
where
\begin{equation}
\hL_{\vartheta}\equiv{1\over\sth}{\partial\over\partial\vartheta}\left(
\sth{\partial\over\partial\vartheta}\right),\quad
\hL_{\varphi}\equiv{1\over\sthsq}
{\partial^2\over\partial\varphi^2}.
\end{equation}
After expanding the unknown function $\dphinow(r,\vartheta,\varphi)$
and the source term $\Smath^{(n)}(r,\vartheta,\varphi)$ in spherical
harmonics (or Fourier-Legendre) components 
\begin{equation}
\dphinow(r,\vartheta,\varphi)=\sum_{l,m}\dphinowlm(r)Y_l^m(\vartheta,\varphi),
\quad
\Smath^{(n)}(r,\vartheta,\varphi)=\sum_{l,m}\Smathlmnow(r)Y_l^m(\vartheta,\varphi),
\label{eqlegendre}
\end{equation}
equation~(\ref{poisson}) takes the simple one-dimensional form
\begin{equation}
{1\over r}\left[{d \over d r}\left({r^2{d
      \over d r}}\right)
-l(l+1)\right]\dphinowlm(r)=r\Smathlmnow(r), 
\label{eqdphilm}
\end{equation}
where we multiplied both sides by $r$ to avoid the singularity in the
source term for the astrophysically relevant case of $\rho\sim r^{-1}$
central density profiles.  Equation~(\ref{eqdphilm}), involving
derivatives only in the radial coordinates, is discretized by central
finite differences.

\subsection{Implementation of the relaxation procedure.}
\label{rsolv}

The relaxation scheme in N-MODY is implemented in the following
steps:
\begin{enumerate}
\item At the initial time $t=0$ in N-body simulations (and in the case
  of static models) the guess solution $\phiini$ is chosen to be the
  spherically symmetric MOND solution of the angle-averaged density
  distribution
\begin{equation}
\rho_{0,0}(r)={1\over 4
  \pi}\int_0^{2\pi}\int_0^{\pi}\rho(r,\vartheta,\varphi)\sin\vartheta
  \,d\vartheta\,d\varphi,
\end{equation}
which is easily derived from the corresponding spherical Newtonian
solution \citep{Bek84}.  In particular this guess solution satisfies
the boundary condition, providing the values of the potential
$\phiini$ and of the radial acceleration $g_r^{(0)}$ at the boundary
grid point $r_{\Nr}$.  At $t>0$ in N-body simulations the guess
solution is provided by the numerical solution found in the previous
timestep.
\item For assigned acceleration $\gvnow(r,\vartheta,\varphi)$ at the
  iteration level $n$, the source term
  $r\Smath^{(n)}(r,\vartheta,\varphi)$ is evaluated, using central
  finite differences to approximate space derivatives.
\item The source term is then transformed into Fourier-Legendre
  components by
\begin{equation}
\Smathlm^{(n)}(r_i)={1\over 4
  \pi}\int_0^{2\pi}e^{-im\varphi}d\varphi\int_0^{\pi}\Smath^{(n)}(r_i,\vartheta,\varphi)
  P_{l,m}(\vartheta)\sin\vartheta d\vartheta
=\sum_j\sum_k{\Smath^{(n)}(r_i,\vartheta_j,\varphi_k)e^{-im\varphi_k}P^{-1}_{l,m}(\vartheta_j)},
\end{equation}
where $P^{-1}_{l,m}(\vartheta_j)$ is the inverse of the matrix
$P_{l,m}(\vartheta_j)$ of the associated Legendre polynomials, so
$\sum_j{P^{
}_{l,m}(\vartheta_j)P^{-1}_{l',m}}(\vartheta_j)=\delta_{l,l'}$.
\item The operator in equation~(\ref{eqdphilm}) is discretized in the
  radial coordinate by using finite differences for first and second
  derivatives.  It results a tri-diagonal or penta-diagonal matrix of
  order $\Nr+1$ that can be easily inverted by using a standard
  Lower-Upper triangular (LU) decomposition to solve for the
  $\dphinowlm$ variables with boundary conditions
  $\dphinowlm(r_{\Nr})=0$.
\item The potential increments $\dphinowlm(r_i)$ are back transformed
  into the $(\vartheta,\varphi)$ coordinate space and the
  corresponding acceleration increments $\delta\gvnow$ are evaluated
  using finite differences, to move to the next level solution
  $\gvplus=\gvnow+\dgvnow$ until convergence is achieved.  Note that
  in this relaxation scheme the potential is never used for
  intermediate solutions, only the potential increments being
  required. The final acceleration field is a potential gradient
  because the guess and all the intermediate solutions keep the
  irrotational form.  The final potential field is computed (when
  needed for outputs) from the final acceleration field by numerical
  inversion of $\nabla \phi= -\gv$.

\item Convergence is achieved when $\Vert \delta g/g\Vert_{\infty} <
  \varepsilon$, where $\delta g=\Vert \delta g \Vert$, $\varepsilon$
  is a tolerance parameter, and $\Vert ...\Vert_{\infty}$ is the maximum
  norm over all grid points.
\end{enumerate}

\section{Particle-mesh scheme and time integration}
\label{secpm}

For a given set of $N$ point particles with mass $m=M/N$, the
Cartesian positions of the particles $(\xp,\yp,\zp),\,p=1,...,N$, are
first converted into spherical coordinates:
\begin{equation}
\rp=\sqrt{\xp^2+\yp^2+\zp^2},\quad\xip=\tan^{-1}(\rp/L)^{1/\alpha},\quad
\vartheta_p=\cos^{-1}(z_p/r_p),\quad\varphi_p=\tan^{-1}(\yp/\xp),
\end{equation}
where the coordinate $\xi$ is defined by equation~(\ref{eqmapping}).
The mass of the particles is deposited on the radial grid, using
linear or quadratic shape functions $S(u-u_p)$ of compact support for
each coordinate $u$. The resulting mass density at the grid point of
indices $(i,j,k)$ is
\begin{equation}
M_{i,j,k}\equiv\left[\rho r^2 r^{\prime}\sin\vartheta\Dxi\Dth\Dph\right]_{i,j,k}=m\sum_p S(\xi_i-\xi_p)S(\vartheta_j-\vartheta_p)S(\varphi_k-\varphi_p),
\end{equation}
where the sum extends only at the particle positions where the shape
function is non zero. Since the mass assignment scheme is conservative,
\begin{equation}
\sum_{i,j,k}M_{i,j,k}=mN=M.
\end{equation}
The inverse operation to assign the grid defined acceleration
components to each particle is performed along similar lines, where
now the same linear or quadratic shape functions act as interpolating
functions. To optimize momentum conservation, we first compute the
components of the derivatives of the potential 
$\gtv=(g_r,rg_{\vartheta},r\sin\vartheta g_{\varphi})$
and then we interpolate them at the particle position $\xx_p$:
\begin{equation}
\gtv(\xx_p)=\sum_{i,j,k}\gtv_{i,j,k}S(\xi_i-\xi_p)S(\vartheta_j-\vartheta_p)
S(\varphi_k-\varphi_p).
\end{equation}
Finally, the interpolated spherical components
\begin{equation}
[g_r]_p=[\gt_r]_p,\quad [g_{\vartheta}]_p=[\gt_{\vartheta}/r]_p,\quad [g_{\varphi}]_p=
[\gt_{\varphi}/r\sin\vartheta]_p
\end{equation}
are combined to get the corresponding Cartesian components of the
particle acceleration
\begin{equation}
g_x=[g_R\cos\varphi-g_{\varphi}\sin\varphi]_p,\quad
g_y=[g_R\sin\varphi+g_{\varphi}\cos\varphi]_p,\quad
g_z=[g_r\cos\vartheta-g_{\vartheta}\sin\vartheta]_p,
\end{equation}
where $g_R=g_r\sin\vartheta+g_{\vartheta}\cos\vartheta$.

In N-MODY time integration is performed with a either second-order or
fourth-order standard leapfrog scheme.  The second-order algorithm is 
made of the following four steps:
\begin{enumerate}
\item half-step position move: $\xx_p(t+\Delta t/2)=\xx_p(t)+\Delta t\,\vv_p(t)/2$;
\item evaluation of the particles acceleration at the current time: $\gv_p(t+\Delta t/2)$;
\item one-step velocity move $\vv_p(t+\Delta t)=\vv_p(t)+\Delta t\,\gv_p(t+\Delta t/2)$;
\item a second half-step position move: $\xx_p(t+\Delta t)=\xx_p(t+\Delta t/2)+\Delta t\,\vv_p(t+\Delta t)/2$.
\end{enumerate}
The implemented fourth order leapfrog scheme is obtained by cycling
three times this basic scheme with timestep $\Delta t$ replaced by
$(c_1,c_2,c_1)\Delta t$ respectively, where $c_1=1/(2-2^{1/3})$ and
$c_2=1-2c_1$.  The timestep, which is the same for all particles, is
adaptive in time, being determined by the leapfrog stability threshold
$\Delta t=\eta/\sqrt{\max|\nabla\cdot\gv|}$, where the maximum is
evaluated over all grid points and $\eta$ is a dimensionless parameter
(typically $\eta=0.3$).

\section{Performances and applications}
\label{parper}

In \cite{Cio06} we tested the potential solver of N-MODY by comparing
the numerical results with analytic MOND potentials of special
axisymmetric and triaxial density distributions.  The N-body code was
tested in \cite{Nip07a}, where we also discussed in detail the
conservation of linear momentum, angular momentum and energy in
N-MODY. Here we give a brief account of the performances of the code
in typical applications.

By choosing a tolerance parameter $\varepsilon=10^{-3}$ for
convergence in maximum norm (corresponding to $\simeq 10^{-4}$ in
r.m.s. norm $\Vert ...\Vert_{\rm rms}$), with a relaxation parameter
$\omega^{-1}=0.3-0.5$, the N-MODY solver provides the required
solution in $5-10$ iterations. For typical N-body systems, these error
bounds correspond to an approximation in maximum norm of
equation~(\ref{eqdiv}), of $\Vert\hM(\phi)\Vert_{\infty}\simeq 0.1$
and $\Vert\hM(\phi)\Vert_{\rm rms}\simeq 0.01$.
\begin{figure}[]
\begin{center}
\resizebox{0.5\hsize}{!}{\includegraphics[clip=true]{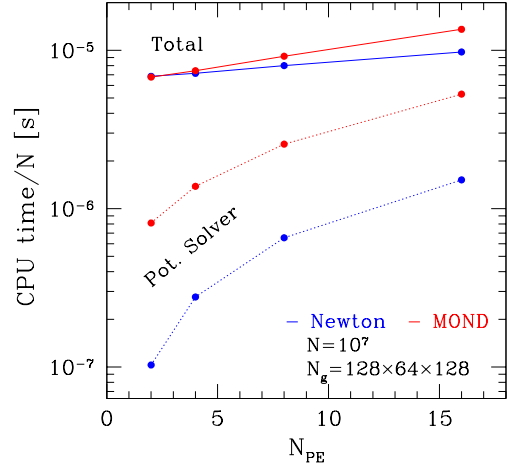}}
\caption{ \footnotesize CPU time (summed over all PEs) per particle
per timestep as a function of the number of PEs for N-MODY simulations
of MOND (red) and Newtonian (blue) \cite{Her90} spheres.}
\end{center}
\label{fig_cpu}
\end{figure}
The computational time needed in the MOND potential solver scales
roughly with the total number of grid points $N_g=(\Nr+1)\Nth\Nph$. In
fact, in the serial version of the code (only one PE used), the MOND
solver needs $\sim 10\%$ of the time spent to run particles, in a
typical configuration where $N\sim10N_g$.  For large grids, and lower
number of particles per cell, some advantages are obtained by
parallelization of the MOND solver. To that purpose, when running with
$\Npe$ processors, N-MODY adopts a domain decomposition during the
iteration cycle, by assigning at each PE only a sector of the
$\vartheta$ coordinate containing $\Nth/\Npe$ grid points
($\Nth\geq4\Npe$).  This simple strategy results to be effective, even
if several operations still require the full grid, resulting in $\sim
70\%$ of parallelization rate.  In Fig.~\ref{fig_cpu} we plot the CPU
time (summed over all PEs) per timestep per particle as a function of
the number of PEs for N-MODY simulations of a \cite{Her90} sphere in
Newtonian gravity (blue) and in MOND \citep[red; with acceleration
parameter $\kappa=1$, see][]{Nip07c}. Both simulations, using $N=10^7$
particles and $N_g=128\times64\times128$ grid points, were run on an
IBM Linux Cluster (Pentium IV/3 GHz PCs).  In the diagram we
distinguish the total CPU time (solid lines) and the CPU time spent by
the potential solver (dotted lines).  In all cases the total CPU time
per particle is $\lsim 10^{-5}$ s, but it is apparent that the
parallelization is not as efficient in MOND as in the case of
Newtonian gravity. This is due to the contribution of the potential
solver, which is only partially parallelized: while in Newtonian
simulations the computational cost of the solver is always negligible,
in MOND simulations it is non-negligible when $\Npe\gsim 10$, because
the iterative procedure requires a factor of 5-10 more time than the
inversion of the Poisson equation.

We have already applied N-MODY to study dissipationless collapse
\citep{Nip07a}, phase mixing \citep{Cio07}, galaxy merging
\citep{Nip07c} and dynamical friction \citep{Nip08} in MOND. We found
that these dynamical processes are profoundly different in MOND and in
Newtonian gravity.  In particular violent relaxation, phase mixing and
merging take significantly longer in MOND than in Newtonian gravity,
while dynamical friction is more effective in a MOND system than in an
equivalent Newtonian system with dark matter.

\smallskip
 N-MODY is publicly available upon request to the authors. It is
written in FORTRAN 90 and can be compiled and run as either a parallel
or a serial code.

\begin{acknowledgements}
The code has been partly developed and tested at CINECA, Bologna, with
CPU time assigned under the INAF-CINECA agreements 2006/2007 and
2007/2008.
\end{acknowledgements}

\bibliographystyle{aa}

\end{document}